\documentclass[twocolumn,preprintnumbers,amsmath,amssymb]{revtex4}

\usepackage{graphicx}
\usepackage{bm}

\begin{document}

\title{Measurement of Rashba and Dresselhaus spin-orbit magnetic fields}

\author{Lorenz Meier$^{1,2}$, Gian Salis$^1$, Ivan Shorubalko$^2$, Emilio Gini$^3$, Silke Sch\"on$^3$, Klaus Ensslin$^2$}
\affiliation{$^1$IBM Research, Zurich Research Laboratory,
S\"aumerstrasse 4, 8803 R\"uschlikon, Switzerland\\
$^2$Solid State Physics Laboratory, ETH Zurich, 8093 Zurich,
Switzerland\\
$^3$FIRST Center for Micro- and Nanosciences, ETH Zurich, 8093
Zurich, Switzerland}

\date{15 July 2007}

\begin{abstract}
Spin-orbit coupling is a manifestation of special relativity. In the
reference frame of a moving electron, electric fields transform into
magnetic fields, which interact with the electron spin and lift the
degeneracy of spin-up and spin-down states. In solid-state systems,
the resulting spin-orbit fields are referred to as Dresselhaus or
Rashba fields, depending on whether the electric fields originate
from bulk or structure inversion asymmetry, respectively. Yet, it
remains a challenge to determine the absolute value of both
contributions in a single sample. Here we show that both fields can
be measured by optically monitoring the angular dependence of the
electrons' spin precession on their direction of movement with
respect to the crystal lattice. Furthermore, we demonstrate spin
resonance induced by the spin-orbit fields. We apply our method to
GaAs/InGaAs quantum-well electrons, but it can be used universally
to characterise spin-orbit interactions in semiconductors,
facilitating the design of spintronic devices.
\end{abstract}

\maketitle

Symmetry-breaking electric fields in semiconductors induce a spin
splitting, because electric fields appear to a moving electron as
magnetic fields, which interact with the electron spin and couple it
with the electron momentum, or wave vector, $\boldsymbol{k}$. In
zinc-blende-type crystals, such as GaAs, the electric fields
resulting from the lack of an inversion centre lead to bulk
inversion asymmetry (BIA) and to the Dresselhaus term in the
Hamiltonian~\cite{Dresselhaus1955}. In the conduction band, its
coupling is linear or cubic in $k$ with proportionality constants
$\beta$ and $\gamma$, respectively. In heterostructures, additional
electric fields are introduced owing to structure inversion
asymmetry (SIA), giving rise to the Rashba term~\cite{Bychkov1984},
which for conduction-band electrons is linear in $k$ with coupling
constant $\alpha$. Both contributions have been extensively
studied~\cite{WinklerBuch}, since a potential use of electron spins
in future devices (e.g. a spin transistor~\cite{Datta1990}) requires
precise control of the spin's environment and of the Dresselhaus and
Rashba fields~\cite{Schliemann2003}. Spin-orbit fields also
contribute to spin decoherence~\cite{DYakonovPerel1971}.

In two-dimensional systems, such as quantum wells (QWs), usually
$\alpha \gg \beta$ and $\gamma \approx 0$~\cite{Lommer1988, Luo1990,
Nitta1997,Schapers1998}. Therefore, measurements of the spin-orbit
coupling initially focused on the Rashba term in QWs and
concentrated on the study of beatings in Shubnikov--de-Haas
oscillations~\cite{Das1989, Luo1990,
Engels1997,Schapers1998,Hu1999}, whose interpretation, however, is
debated~\cite{Pfeffer1999, Brosig1999}. More recent experiments
include the investigation of antilocalization in
magnetotransport~\cite{Koga2002} or the analysis of
photocurrents~\cite{Ganichev2004}. In the latter experiment, the
ratio $\alpha/\beta$ could be determined. A gate-induced transition
from weak localization to antilocalization allowed the
discrimination between Rashba, as well as linear and cubic
Dresselhaus contributions to the spin-orbit field~\cite{Miller2003}.
Tuning of the Rashba coupling has been achieved by introducing
additional electric fields from gates~\cite{Nitta1997, Grundler2000}
or by changing the electron density~\cite{Heida1998, Matsuyama2000}.

The influence of effective spin-orbit magnetic fields on optical
measurements in a heterostructure was already measured in
1990~\cite{Kalevich1990}, and the spin-orbit-induced precession of
spin packets was observed more than a decade
later~\cite{KatoNature2004, Crooker2005}. Remarkably, the in-plane
spin-orbit fields in a QW can lead to an out-of-plane spin
polarization~\cite{KatoPRL2004}. In ref.~\cite{EngelPRL2007}, it was
pointed out that although spin-orbit and external magnetic fields
can be added to describe spin precession~\cite{Kalevich1990}, a more
complicated concept has to be evoked when accounting for the
generation of an out-of-plane spin polarization.

In this Article, we show that both Rashba and Dresselhaus fields in
the conduction band can be determined by measuring the spin
precession of optically polarized electron spins as a function of
the direction of their drift momentum. We find good agreement with
the assumption that the spins precess about the sum of the effective
spin-orbit and an external magnetic field $\textbf{B}_0$. In
Sect.~\ref{sec1}, we derive an expression for the magnitude of this
total magnetic field $B_\textrm{tot}$ as a function of the angles
$\theta$ and $\varphi$ that $\textbf{B}_0$ and the electron drift
momentum $\hbar\boldsymbol{k}$ include with the crystal's
$[1\overline{1}0]$ axis. These predictions provide a good
description of the experimental data
$B_\textrm{tot}(\theta,\varphi)$ of three (001)~GaAs/InGaAs~QW
samples given in Sect.~\ref{sec2}, and allow us to separately
determine the Rashba and Dresselhaus contributions to the spin-orbit
field. In Sect.~\ref{sec3}, we demonstrate spin resonance induced by
oscillating spin-orbit fields.

\section{Theoretical expectations}\label{sec1}
\begin{figure}[tb]
\includegraphics[width=90mm]{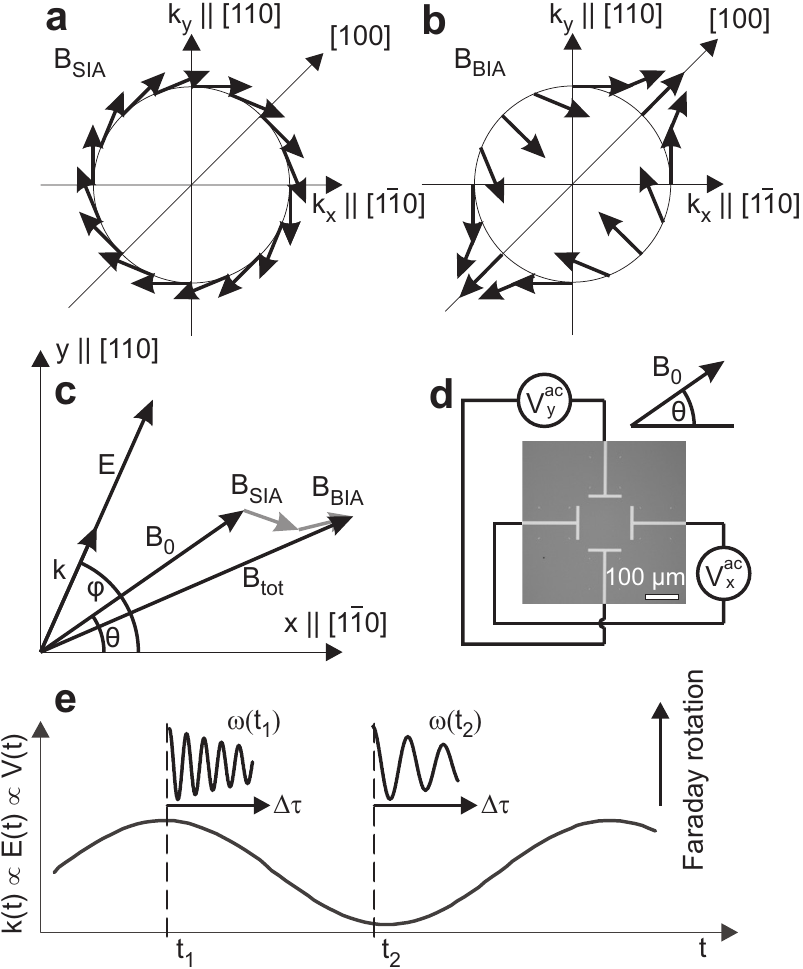}
\caption{\label{fig:fig1} Orientation of the magnetic and electric
fields and measurement setup. \textbf{a,} Rashba and \textbf{b,}
Dresselhaus spin-orbit fields for different orientations of the
$\boldsymbol{k}$-vector on a unit circle. \textbf{c,} Electric and
magnetic fields involved in the experiment. \textbf{d,} Optical
microscopy image of the sample and wiring of the gates. \textbf{e,}
The electron precession frequency $\omega(t)$ is determined at
different phases of the oscillating wave vector $k(t)$.}
\end{figure}
Neglecting cubic terms, the Rashba and Dresselhaus spin-orbit
couplings in a QW are linear in $\boldsymbol{k}$ and can be
described by an effective magnetic
field~\cite{GanichevReview2003,WinklerBuch},
 \begin{equation}\label{eq:SIABIA}
    \textbf{B}_{\textrm{SIA}} = \frac{\alpha}{g \mu_B} \binom{k_y}{-k_x}, \quad \textrm{and}
    \quad
    \textbf{B}_{\textrm{BIA}} = \frac{\beta}{g \mu_B} \binom{k_y}{k_x},
 \end{equation}
for a coordinate system with base vectors $\hat{\textbf{x}}\parallel
[1\overline{1}0]$ and $\hat{\textbf{y}}\parallel [110]$. Here, $g$
is the electron's $g$-factor and $\mu_B$ the Bohr magneton. Both
fields are in the plane of the QW, but while
$\textbf{B}_{\textrm{SIA}}$ is always perpendicular to
$\boldsymbol{k}$ (Fig.~\ref{fig:fig1}{a}),
$\textbf{B}_{\textrm{BIA}}$ has a different geometrical dependence
on $\boldsymbol{k}$ (Fig.~\ref{fig:fig1}{b}). A conduction-band
electron experiences the total magnetic field
$\textbf{B}_{\textrm{tot}} = \textbf{B}_0 +
\textbf{B}_{\textrm{SIA}} + \textbf{B}_{\textrm{BIA}}$
(Fig.~\ref{fig:fig1}{c}). With a time-dependent $\boldsymbol{k}(t) =
k_0\sin{(2\pi\nu t)}\cdot\left(\cos{\varphi},\sin{\varphi}\right)$,
we obtain for the total magnetic field square
\begin{equation}\label{eq:totfieldmagnitude}
\begin{split}
    B_\textrm{tot}^2 & = B_0^2
    \times \\
    ( 1 & + \frac{2k_0\sin{(2\pi\nu t)}}{g
    \mu_B B_0}\left([\alpha+\beta]\cos{\theta}\sin{\varphi} +
    [\beta-\alpha]\sin{\theta}\cos{\varphi}\right)\\
    & + \left( \frac{k_0\sin{(2\pi\nu t)}}{g \mu_B B_0} \right)^2
    \left( \alpha^2 + \beta^2 - 2\alpha\beta\cos{2\varphi} \right)
    ).
\end{split}
\end{equation}

$B_\textrm{tot}^2(t)$ is expected to oscillate around $B_0^2$ with
frequencies $\nu$ and $2\nu$. If cubic Dresselhaus terms were
included in $B_\textrm{tot}$, additional terms proportional to
$k^n$, $n=3,4,6$ would appear in Eq.~(\ref{eq:totfieldmagnitude})
and induce oscillations at frequencies $n\nu$. Assuming that
$|B_\textrm{SIA}|, |B_\textrm{BIA}| \ll B_0$, we expand the square
root of Eq.~(\ref{eq:totfieldmagnitude}) up to second order in
$k_0/B_0$, and obtain, using Eq.~(\ref{eq:SIABIA}),
\begin{equation}\label{eq:RootExpanded}
    B_\textrm{tot}(t) \approx B_0 + A(\theta,\varphi)\sin{(2\pi\nu t)} + B(\theta,\varphi)\sin^2{(2\pi \nu
    t)},
\end{equation}
with
\begin{equation*}\label{eq:AB}
\begin{split}
  A(\theta,\varphi)  = \,&\left(B_\textrm{BIA} + B_\textrm{SIA}\right)\cos{\theta}\sin{\varphi}\\
  + &\left(B_\textrm{BIA} - B_\textrm{SIA}\right)\sin{\theta}\cos{\varphi}\textrm{, \quad and}\\
  B(\theta,\varphi) = [&\left(B_\textrm{BIA} + B_\textrm{SIA}\right)\sin{\theta}\sin{\varphi}\\
  - &\left(B_\textrm{BIA}- B_\textrm{SIA}\right)\cos{\theta}\cos{\varphi}]^2/B_0.
\end{split}
\end{equation*}

By measuring the oscillation amplitude of $B_\textrm{tot}(t)$ for
varying angles $\theta$ and $\varphi$, we can extract the Rashba and
Dresselhaus contributions to the spin-orbit magnetic field.

\section{Experiments}\label{sec2}
To induce an oscillating spin-orbit field, we impose an oscillating
drift momentum $\hbar\boldsymbol{k}(t)$ on the QW electrons by
applying an in-plane a.c.\ electric field $\textbf{E}(t) =
\textbf{E}_0\sin{(2\pi\nu t)}$, $\nu = 160$~MHz, at an angle
$\varphi$ with the $x$-axis, see Fig.~\ref{fig:fig1}{c}. In the
diffusive limit, electron scattering occurs fast on the timescale
$1/\nu$, and $\boldsymbol{k}(t) \propto \textbf{E}(t)$ (see
Methods). We monitor the spin precession frequency $\omega(t) =
g\mu_B B_\textrm{tot}(t)/\hbar$ of optically polarized electron
spins at different times $t$ (see Fig.~\ref{fig:fig1}{e}) using
time-resolved Faraday rotation (TRFR, see Methods).

\begin{figure}[tb]
\includegraphics[width=90mm]{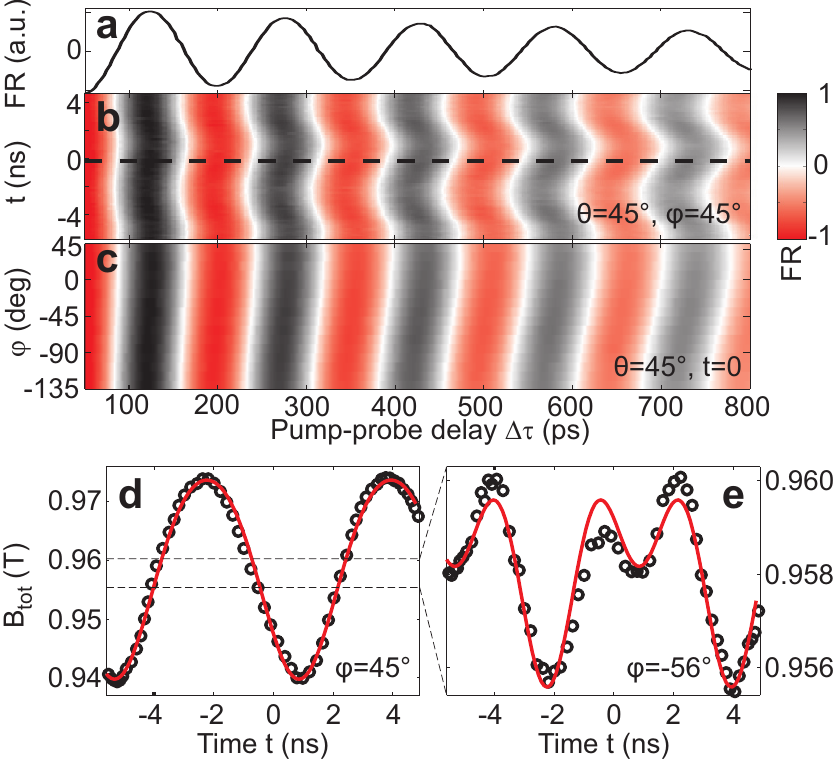}
\caption{\label{fig:fig2} TRFR signal measured at different times
$t$ and electric field angles $\varphi$, at $\theta = 45^\circ$.
\textbf{a,} Faraday rotation vs. pump-probe delay. \textbf{b,} TRFR
scans at different times $t$, the dashed line indicates $t=0$.
\textbf{c,} TRFR scans at different angles $\varphi$, at $t=0$.
\textbf{d, e,} Total magnetic field as a function of $t$ for
\textbf{d,} $\varphi = 45^\circ$ and \textbf{e,} $\varphi =
-56^\circ$. The solid line is a fit to Eq.~(\ref{eq:RootExpanded}).}
\end{figure}

Figure~\ref{fig:fig2}{b} shows TRFR oscillations at different $t$.
Owing to the oscillating spin-orbit field, $\omega(t)$ and
consequently $B_\textrm{tot}(t)$ change periodically with $t$.
Likewise, $\omega(t)$ changes with the angle $\varphi$
(Fig.~\ref{fig:fig2}{c}), as predicted by
Eq.~(\ref{eq:RootExpanded}). The fit to this equation with $B_0$,
$A(\theta,\varphi)$, and $B(\theta,\varphi)$ as fit parameters
matches the data points very well (Fig.~\ref{fig:fig2}{d}), with
$B_0 = 0.958$~T, in agreement with Hall probe measurements of the
external magnetic field. For most $\varphi$, we find $A \gg B$. The
quadratic term in $k(t)$, $B(\theta,\varphi)$, which contributes to
oscillations with frequency $2\nu$, is visible in the experiment
only when rotating \textbf{E} to an angle $\varphi$ at which the
$k$-linear term $A(\theta,\varphi)$ is weak, see
Fig.~\ref{fig:fig2}{e}. Apart from the geometrical dependence, the
amplitude of $B(\theta,\varphi)$ is suppressed by a factor
$(|B_\textrm{BIA}|+|B_\textrm{SIA}|)/B_0 \approx 0.03$, i.e.\ by
more than one order of magnitude compared with $A(\theta,\varphi)$.
Higher-order contributions were below the detection limit (roughly
1/4 of the second-order effects), which indicates that in our
samples, \emph{cubic} Dresselhaus terms are not relevant.

\begin{figure}[tb]
\includegraphics[width=90mm]{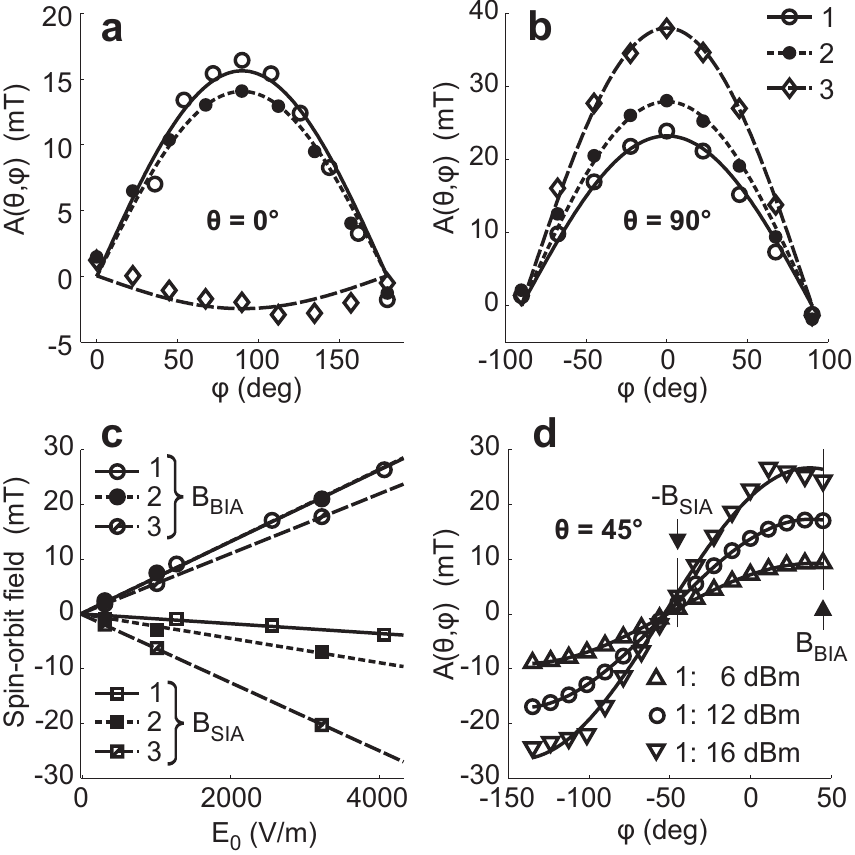}
\caption{\label{fig:fig3}Spin-orbit fields as a function of
$\varphi$ for $\theta = 0, 90^\circ$ and $45^\circ$. Depending on
$\theta$, we measure \textbf{a,} $(B_\textrm{BIA} +
B_\textrm{SIA})\sin{\varphi}$ [at $\theta = 0$] or \textbf{b,}
$(B_\textrm{BIA} - B_\textrm{SIA})\cos{\varphi}$ [at $\theta =
90^\circ$]. The gate modulation amplitude was 12~dBm for sample~1
and 14~dBm for samples~2 and 3. \textbf{c,} Both $B_\textrm{SIA}$
and $B_\textrm{BIA}$ increase linearly as a function of the electric
field. \textbf{d,} For $\theta = 45^\circ$, the measured spin-orbit
field is $B_\textrm{BIA}\cos{(\varphi - \pi/4)} +
B_\textrm{SIA}\sin{(\varphi - \pi/4)}$ with a linear dependence on
the applied gate voltage. At $\varphi = 45^\circ$, we directly
measure $B_\textrm{BIA}$, and at $\varphi = -45^\circ$,
$-B_\textrm{SIA}$.}
\end{figure}

Given that already second-order terms are strongly suppressed, we
restrict our analysis to the linear term $A(\theta,\varphi)$. For
$\theta = 0$ and $90^\circ$, $A(\theta, \varphi)$ is given by
$(B_\textrm{SIA}+B_\textrm{BIA})\sin{\varphi}$ and
$(B_\textrm{BIA}-B_\textrm{SIA})\cos{\varphi}$, respectively. This
dependence is observed in the experiment, as shown in
Fig.~\ref{fig:fig3}{a and b}. The measured data points clearly
follow $\sin{\varphi}$ for $\theta = 0$ and $\cos{\varphi}$ for
$\theta = 90^\circ$ (solid lines). From the two measurements at
$\theta = 0$ and $90^\circ$, we can extract the spin-orbit magnetic
fields $B_\textrm{SIA}$ and $B_\textrm{BIA}$. Normalized to a gate
modulation amplitude of $V_0 = 2$~V ($\approx 13$~dBm),
corresponding to an electric field of $E_0 \approx 2,900$~V/m, we
find $B_\textrm{SIA} = -4.2, -8.5$, and $-17.6$~mT and
$B_\textrm{BIA} = 21.6, 21.1$, and $ 15.7$~mT for samples~1, 2, and
3, respectively. Note that as $t$ is known up to an offset $t_0$,
the sign of $A(\theta,\varphi)$ is arbitrary, leading to an
uncertainty in the absolute sign of $B_\textrm{SIA}$ and
$B_\textrm{BIA}$ (the relative sign is obtained). We choose
$B_\textrm{BIA}> 0$. As a function of the magnitude of the applied
electric field $E_0$, $B_\textrm{SIA}$ and $B_\textrm{BIA}$ increase
linearly (see Fig.~\ref{fig:fig3}{c}), as expected from the linear
relation between $k(t)$ and $E(t)$ and Eq.~(\ref{eq:SIABIA}). We
have conducted the same measurements at different magnitudes of $B_0
=$~0.55 and 0.82~T, and found similar values for the spin-orbit
fields.

As discussed above, measurements at two angles, $\theta = 0$ and
$90^\circ$, were needed to obtain $B_\textrm{SIA}$ and
$B_\textrm{BIA}$. At $\theta = 45^\circ$ both $B_\textrm{SIA}$ and
$B_\textrm{BIA}$ can be determined simultaneously. This is because
not only the amplitude, but also the phase of the oscillation in
$\varphi$ contains information about the spin-orbit fields. The
zero-crossing occurs at $\varphi_0$ =
$\arctan{[(B_\textrm{SIA}-B_\textrm{BIA})/(B_\textrm{SIA}+B_\textrm{BIA})]}$,
compared with $\varphi_0 = \theta$ for $\theta = 0$ and $90^\circ$.
For $\theta=\varphi=0$ and $90^\circ$, $\textbf{B}_\textrm{SIA}$ and
$\textbf{B}_\textrm{BIA}$ are perpendicular to $\textbf{B}_0$
(Fig.~\ref{fig:fig1}{a and b}) and $A(\theta,\varphi)$ vanishes,
because it is equal to the component of $\textbf{B}_\textrm{SIA} +
\textbf{B}_\textrm{BIA}$ along the direction of $\textbf{B}_0$. If
however $\theta=\varphi=45^\circ$, $\textbf{B}_\textrm{SIA}$ still
is perpendicular to $\textbf{B}_0$, but $\textbf{B}_\textrm{BIA}$ is
now parallel, and $A(\theta,\varphi) = B_\textrm{BIA}$.

The measurement at $\theta = 45^\circ$ is shown in
Fig.~\ref{fig:fig3}{d}, with a fit to Eq.~(\ref{eq:RootExpanded}).
For $V_0 = 2$~V, we extract the spin-orbit fields $B_\textrm{SIA} =
-2.4$~mT and $B_\textrm{BIA} = 19.1$~mT for sample~1. These values
correspond well to the values obtained from $\theta = 0$ and
$90^\circ$. Relative variations in $B_\textrm{SIA}$ of up to 50\%
(but far less in $B_\textrm{BIA}$) occurred for different cool-downs
of the same sample, which we attribute to the freezing of electron
states in the QW interface or to strain.

Knowing the electron $g$-factor and drift wave vector $k$, we can
calculate the coupling constants $\alpha$ and $\beta$ from
$B_\textrm{SIA}$ and $B_\textrm{BIA}$ using Eq.~(\ref{eq:SIABIA}).
For sample~3, the mobility is known (see Methods), and with a
numerical simulation of $E_0$, we obtain $\alpha = \hbar g \mu_B
B_\textrm{SIA}/m^\star \mu E_0 = 1.5\times 10^{-13}$~eV$\cdot$m and
$\beta = \hbar g \mu_B B_\textrm{BIA}/m^\star \mu E_0 = -1.4\times
10^{-13}$~eV$\cdot$m, where we have used $g=-0.27$, as independently
measured by TRFR in a known external magnetic field, and assuming
$g<0$. Previous experiments report $\alpha \approx 5-10 \times
10^{-12}$~eV$\cdot$m on
In$_{0.53}$Ga$_{0.47}$As/In$_{0.52}$Al$_{0.48}$As QWs or
heterostructures~\cite{Nitta1997, Hu1999, Koga2002} and in InAs/AlSb
QWs~\cite{Heida1998} and assume $\alpha \gg \beta$. The Rashba
coupling is proportional to the average electric field in the
\emph{valence} band~\cite{WinklerBuch}, including contributions from
band discontinuities. We estimate the valence band offset in our QWs
to be on the order of 10~meV, which is much smaller than in those
previously investigated structures, and explains our small value of
$\alpha$. Our $\alpha$ is about four times larger than that reported
in ref.~\cite{KatoNature2004}, where an $\textrm{In}_{0.07}$GaAs
epilayer (10 times thicker than our QW) was studied. There, the
interfaces play a minor role and strain-induced spin-orbit coupling
predominates. The linear Dresselhaus term is expected to scale with
the extent of the wave function in the confinement direction,
$\langle k_z^2 \rangle$. For an infinitely deep well with width
$\ell$, $k_z \propto 1/\ell$, and $\beta \propto 1/\ell^2$. Assuming
that samples~1 and 2 have similar mobilities, we observed almost the
same $\beta$, even though the QW in sample~2 is twice as wide as
that in sample~1. This could be attributed to inhomogeneous In
deposition during growth, leading to a triangular confinement
potential, where the nominal QW width has less influence on $\beta$.

\section{Electric-dipole-induced spin resonance}\label{sec3}
\begin{figure}[tb]
\includegraphics[width=90mm]{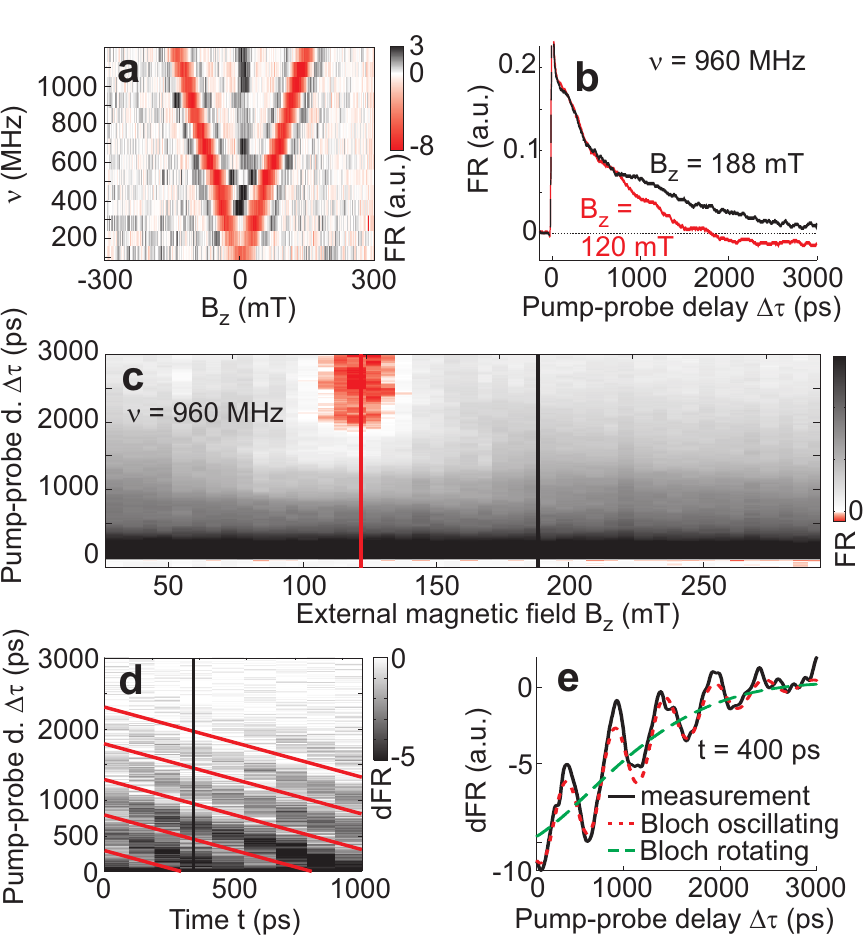}
\caption{Spin resonance induced by an oscillating spin-orbit field.
\label{fig:fig4}\textbf{a,} TRFR signal at $\Delta\tau = 3$~ns for
varying external magnetic fields $B_z$ and electric field
frequencies $\nu$. Resonance is observed with $|g| = 0.57$.
\textbf{b,} TRFR scans on (red line) and off (black line) resonance.
\textbf{c,} TRFR scans $\Theta_F(\Delta\tau)$ at different $B_z$. On
resonance, the spins precess coherently about the spin-orbit
induced-tipping field. \textbf{d,} Differential TRFR signal
$\dot{\Theta}_F(\Delta\tau + t)$ on resonance ($\nu = 960$~MHz), red
lines are guides to the eye at $\Delta\tau = t$. \textbf{e,}
Measured $\dot{\Theta}_F(\Delta\tau) \propto \dot{S_z}(\Delta\tau)$
(solid line) and Bloch simulations with linearly oscillating
(dotted) and rotating tipping field (dashed), at $\nu = 960$~MHz and
$t=400$~ps (solid line in \textbf{d}).}
\end{figure}

In electron spin resonance (ESR) experiments, spins that are
initially polarized along the direction of a static magnetic field
$B_0 = B_z$ perform Rabi oscillations between the states parallel
and anti-parallel to $B_z$ if an a.c.\ magnetic field (the tipping
field) is applied in the plane perpendicular to $B_z$ and at the
Larmor frequency $\nu = g\mu_B B_z/h$. Instead of an a.c.\ magnetic
field, we use an a.c.\ electric field $E_x(t)$ in the plane of the
QW. It induces an oscillating spin-orbit field $B_y(t)$, which can
serve as a tipping field for ESR, in this context referred to as
electric-dipole-induced spin resonance (EDSR)~\cite{Duckheim2006}.
The measurements presented in Fig.~\ref{fig:fig4} have been
conducted in Faraday geometry with sample~1. Here, the external
magnetic field $B_z$ is parallel to the laser propagation and
perpendicular to the QW plane. The pump laser pulse polarizes the
spins into an eigenstate, in line with $B_z$, and the probe pulse
monitors the spin polarization along $z$. In Fig.~\ref{fig:fig4}{a},
the pump-probe delay $\Delta\tau$ has been set to 3~ns, and the
Faraday signal is recorded while sweeping the frequency $\nu$ of
$E_x(t)$ and $B_z$. On resonance, the optically generated spin
polarization precesses about the spin-orbit-induced tipping field,
and the TRFR signal at $\Delta\tau = 3$~ns becomes negative. We
observe spin resonance with $|g| = 0.57$, which is in agreement with
the observed spin precession in Sect.~\ref{sec2}.

In Fig.~\ref{fig:fig4}{b and c}, TRFR scans are collected for
varying $B_z$, monitoring the spin dynamics. At $B_z = 120$~mT, the
Larmor frequency matches the electric field frequency $\nu =
960$~MHz and resonance occurs. Note that the short spin relaxation
time of $\approx 1$~ns strongly reduces the signal. At $\Delta\tau_s
\approx 1800$~ps, the spins have performed a $\pi/2$ Rabi
oscillation, yielding an estimate of the tipping field amplitude
$B_y = 2h/g\mu_B4\Delta\tau_s \approx 35$~mT. Here, the factor 2
takes into account the linearly (and not circularly) oscillating
tipping field~\cite{Bloch1940}. This value agrees well with the
measurements of $|\textbf{B}_\textrm{SIA} - \textbf{B}_\textrm{BIA}|
\approx 33$~mT at a gate modulation amplitude of $V_0 \approx 2.5$~V
(15~dBm) and $\varphi = 0$.

In ESR, one assumes a tipping field that oscillates circularly in
the $x/y$ plane, resulting in a monotonous decrease of the spin
polarization along $z$ during the first $\pi/2$ Rabi oscillation.
The spin dynamics can then be solved analytically in the ``rotating
frame''. In EDSR, the Rabi oscillation on resonance is not steady
with time (Fig.~\ref{fig:fig4}{b}), because the tipping field
oscillates linearly on the $y$-axis instead. The precession of a
spin is described by the Bloch equations (neglecting spin
relaxation)
\begin{equation}\label{eq:Bloch}
  \dot{\textbf{S}} = \frac{g\mu_B}{\hbar}\textbf{B} \times
  \textbf{S},
\end{equation}
from which we find $\dot{S_z}(t') = g\mu_B B_y(t') S_x(t')/\hbar$,
where the dot denotes the time derivative and $t' = \Delta\tau + t$.
The tipping field $B_y(t') \propto \sin{(2\pi\nu t')}$ and with it
$\dot{S_z}(t')$ vanish twice per electric field period $1/\nu$,
resulting in a stepwise decrease of $S_z$ (on resonance, $S_x(t')$
and $B_y(t')$ vanish simultaneously). This is shown in
Fig.~\ref{fig:fig4}{d}, where the time derivative of the Faraday
signal $\dot{\Theta}_F(t') \propto \dot{S_z}(t')$ is plotted for
different $\Delta\tau$ and $t$ and for $B_z$ on resonance. Apart
from decaying with time, it is periodic in both $\Delta\tau$ and
$t$, with period $1/2\nu$. The stepwise decrease of $S_z$ can be
reproduced by a numerical solution of the Bloch equations including
a spin-relaxation term with $T_1 = T_2 = 1$~ns, $B_z = 120$~mT, $B_x
= 0$ and $B_y(t) = 34~\textrm{mT}\sin{(2\pi\nu t)}$. We show
$\dot{S_z}$ of this solution in Fig.~\ref{fig:fig4}{e}, together
with the experimentally measured $\dot{\Theta}_F(t')$ and the
corresponding solution with a rotating tipping field. For the
latter, the tipping field has to be reduced by a factor
2~\cite{Bloch1940}.

The technique presented here to unambiguously determine Rashba and
Dresselhaus spin-orbit fields with high precision can be extended to
any semiconductor sample, if optical access to electron spin
precession is provided. This is useful for the optimization of
semiconductor materials and QW designs with increased spin-orbit
fields that can be used for efficient EDSR-based spin manipulation.
Moreover, it might facilitate the tuning of systems where Rashba and
Dresselhaus terms cancel out each other, opening an avenue to study
spin dynamics in this interesting regime.

\section{methods}
\subsection{Application of the electric field}
In the centre of four top-gate electrodes, which enclose a square
with 150~$\mu$m side length (see Fig.~\ref{fig:fig1}{d}), the angle
$\varphi$ of the oscillating electric field $\textbf{E}(t)$ is
determined by the amplitudes $E_x$ and $E_y$ of two superposed
fields along $\hat{\textbf{x}}$ and $\hat{\textbf{y}}$,
$\textbf{E}_0 = E_x\hat{\textbf{x}} + E_y\hat{\textbf{y}}$. $E_x$
and $E_y$ are generated by two phase-locked oscillators, each
driving two opposite electrodes. In the diffusive regime, the
scattering time of the electrons in the QW ($\approx 0.5$~ps from
mobility measurements) is much smaller than $1/\nu$ and, therefore,
their average drift wave vector points along $\textbf{E}(t)$ and its
magnitude is given by $k(t) = m^\star\mu E(t)/\hbar = m^\star\mu
E_0\sin{(2 \pi \nu t)}/\hbar$, with $m^\star$ the electron effective
mass, $\mu$ the electron mobility in the QW, and $\hbar$ Planck's
constant.

\subsection{Time-resolved Faraday rotation} To determine the total
magnetic field acting on the QW electrons in the centre of the four
top-gate electrodes (see Fig.~\ref{fig:fig1}{d}), we employ
time-resolved Faraday rotation (TRFR)~\cite{Crooker1995}. A first,
circularly polarized laser pulse ($P = 400$~$\mu$W, focus diameter
$\approx 15$~$\mu$m, pulse width $\approx 3$~ps) tuned to the
absorption edge of the QW (870~nm) creates a spin polarization
perpendicular to the QW plane. The linear polarization of a second
laser pulse ($P = 50$~$\mu$W), which is transmitted through the
sample at a time delay $\Delta\tau$ with respect to the first pulse,
is rotated by an angle $\Theta_F$ proportional to the spin
polarization along the QW normal. As the spins precess about a local
in-plane magnetic field $\textbf{B}_{\textrm{tot}}$, an oscillating
signal $\Theta_F(\Delta\tau) = \Theta_0 \cos{(\omega \Delta\tau)}
e^{-\Delta\tau/T_2^\star}$ is measured (see Fig.~\ref{fig:fig2}{a}).
The exponential accounts for the finite spin lifetime $T_2^\star$,
and $\omega = g\mu_B B_\textrm{tot}/\hbar$ is proportional to the
magnitude of the total magnetic field $\textbf{B}_{\textrm{tot}}$.
Experiments are performed at $T=40$~K, where effects of nuclear
polarization are negligible~\cite{MeierAPL06}. To probe the spin
precession at a given time $t$ and thus at a given phase of $k(t)
\propto E(t)$, the pulsed laser is phase-locked to the oscillatory
field $E(t)$ with a variable phase shift. We probe the spin
precession during an interval $\Delta\tau =$~0\ldots700~ps, which is
much shorter than the a.c.\ electric field period of $1/\nu =
6,250$~ps. Therefore, $\textbf{E}(t)$ is roughly constant over the
spin precession observed and a well-defined precession frequency
$\omega(t)$ can be obtained, see Fig.~\ref{fig:fig1}{e}. For
experimental reasons, the time $t$ is known up to an offset $t_0$,
which is constant throughout the experiment.

\subsection{Sample structure} Sample 1 is a 20-nm-wide GaAs/InGaAs
QW with an In content of 8.5\%, capped by 21~nm GaAs and grown on a
GaAs substrate by metal organic chemical vapor deposition. Both cap
and well are $n$-doped to maximize the spin
lifetime~\cite{Kikkawa1998}. Sample 2 is similar to sample~1, but
with a QW width of 43~nm. Sample 3 is a GaAs/InGaAs QW with an In
content aimed at 10\%, grown by molecular beam epitaxy, $n$-doped on
both sides and in the 20-nm-wide QW, and capped by 30~nm GaAs. For
this sample, we determined a carrier density $n_s = 5.8 \times
10^{11}$~cm$^{-2}$ and a mobility $\mu = 10,600$~cm$^2$/Vs in a
Van-der-Pauw Hall measurement. Transport measurements of samples~1
and 2 were dominated by strongly localized states, presumably due to
a parallel conductivity from the doping layer, and rendered a
determination of $n_s$ and $\mu$ in the QW impossible. In the
optical measurements, however, the QW could be probed independently
of the doping layer. The fitting parameter $B_0$ in
Eq.~(\ref{eq:RootExpanded}) is independent of $\theta$ for all three
samples, which indicates that the $g$-factor is isotropic in the QW
plane, as expected for (001) GaAs/InGaAs QWs. Top-gate electrodes
are fabricated by evaporating 80~nm Au on a PMMA mask defined by
standard electron-beam lithography techniques.

\section{Addendum}
\subsection{Correspondence}
Correspondence and requests for materials should be addressed to
L.M. (meier.lorenz@phys.ethz.ch) or G.S. (gsa@zurich.ibm.com).

\subsection{Acknowledgements}
The authors wish to acknowledge R.~Leturcq for help with sample
preparation and M.~Duckheim, D.~Loss, R.~Allenspach and T.~Ihn for
discussions. This work was supported by the Swiss National Science
Foundation (NCCR Nanoscale Science).

\subsection{Author contributions}
L.M. performed the experiments and analysed the data in close
collaboration with G.S. Samples were fabricated by L.M. and I.S.,
and grown by E.G. (samples 1 and 2) and S.S. (sample 3). K.E.
initiated the collaboration and supported the project in
discussions.

\subsection{Competing Interests}
The authors declare that they have no competing financial interests.


\begin{thebibliography}{10}
\expandafter\ifx\csname url\endcsname\relax
  \def\url#1{\texttt{#1}}\fi
\expandafter\ifx\csname urlprefix\endcsname\relax\def\urlprefix{URL
}\fi \providecommand{\bibinfo}[2]{#2}
\providecommand{\eprint}[2][]{\url{#2}}

\bibitem{Dresselhaus1955}
\bibinfo{author}{Dresselhaus, G.}
\newblock \bibinfo{title}{Spin-orbit coupling effects in zinc blende
  structures}.
\newblock \emph{\bibinfo{journal}{Phys. Rev.}} \textbf{\bibinfo{volume}{100}},
  \bibinfo{pages}{580--586} (\bibinfo{year}{1955}).

\bibitem{Bychkov1984}
\bibinfo{author}{Bychkov, Y.~A.} \& \bibinfo{author}{Rashba, E.~I.}
\newblock \bibinfo{title}{Oscillatory effects and the magnetic susceptibility
  of carriers in inversion layers}.
\newblock \emph{\bibinfo{journal}{J. Phys. C}} \textbf{\bibinfo{volume}{17}},
  \bibinfo{pages}{6039--6045} (\bibinfo{year}{1984}).

\bibitem{WinklerBuch}
\bibinfo{author}{Winkler, R.}
\newblock \emph{\bibinfo{title}{Spin-Orbit Coupling Effects in Two-Dimensional
  Electron and Hole Systems}}, vol. \bibinfo{volume}{191/2003} of
  \emph{\bibinfo{series}{Springer Tracts in Modern Physics}}
  (\bibinfo{publisher}{Springer, Berlin}, \bibinfo{year}{2003}).

\bibitem{Datta1990}
\bibinfo{author}{Datta, S.} \& \bibinfo{author}{Das, B.}
\newblock \bibinfo{title}{Electronic analog of the electro-optic modulator}.
\newblock \emph{\bibinfo{journal}{Appl. Phys. Lett.}}
  \textbf{\bibinfo{volume}{56}}, \bibinfo{pages}{665--667}
  (\bibinfo{year}{1990}).

\bibitem{Schliemann2003}
\bibinfo{author}{Schliemann, J.}, \bibinfo{author}{Egues, J.~C.} \&
  \bibinfo{author}{Loss, D.}
\newblock \bibinfo{title}{Nonballistic spin-field-effect transistor}.
\newblock \emph{\bibinfo{journal}{Phys. Rev. Lett.}}
  \textbf{\bibinfo{volume}{90}}, \bibinfo{pages}{146801}
  (\bibinfo{year}{2003}).

\bibitem{DYakonovPerel1971}
\bibinfo{author}{D'Yakonov, M.~I.} \& \bibinfo{author}{Perel', V.~I.}
\newblock \bibinfo{title}{Spin relaxation of conduction electrons in
  noncentrosymetric semiconductors}.
\newblock \emph{\bibinfo{journal}{Sov. Phys. Solid State}}
  \textbf{\bibinfo{volume}{13}}, \bibinfo{pages}{3023--3026}
  (\bibinfo{year}{1971}).

\bibitem{Lommer1988}
\bibinfo{author}{Lommer, G.}, \bibinfo{author}{Malcher, F.} \&
  \bibinfo{author}{Rossler, U.}
\newblock \bibinfo{title}{Spin splitting in semiconductor heterostructures for
  B $\rightarrow$ 0}.
\newblock \emph{\bibinfo{journal}{Phys. Rev. Lett.}}
  \textbf{\bibinfo{volume}{60}}, \bibinfo{pages}{728--731}
  (\bibinfo{year}{1988}).

\bibitem{Luo1990}
\bibinfo{author}{Luo, J.}, \bibinfo{author}{Munekata, H.},
  \bibinfo{author}{Fang, F.~F.} \& \bibinfo{author}{Stiles, P.~J.}
\newblock \bibinfo{title}{Effects of inversion asymmetry on electron energy
  band structures in GaSb/InAs/GaSb quantum wells}.
\newblock \emph{\bibinfo{journal}{Phys. Rev. B}} \textbf{\bibinfo{volume}{41}},
  \bibinfo{pages}{7685--7693} (\bibinfo{year}{1990}).

\bibitem{Nitta1997}
\bibinfo{author}{Nitta, J.}, \bibinfo{author}{Akazaki, T.},
  \bibinfo{author}{Takayanagi, H.} \& \bibinfo{author}{Enoki, T.}
\newblock \bibinfo{title}{Gate control of spin-orbit interaction in an inverted
  In$_{0.53}$Ga$_{0.47}$As/In$_{0.52}$Al$_{0.48}$As heterostructure}.
\newblock \emph{\bibinfo{journal}{Phys. Rev. Lett.}}
  \textbf{\bibinfo{volume}{78}}, \bibinfo{pages}{1335--1338}
  (\bibinfo{year}{1997}).

\bibitem{Schapers1998}
\bibinfo{author}{Schapers, T.} \emph{et~al.}
\newblock \bibinfo{title}{Effect of the heterointerface on the spin splitting
  in modulation doped In$_x$Ga$_{1-x}$As/InP quantum wells for B $\rightarrow$
  0}.
\newblock \emph{\bibinfo{journal}{J. Appl. Phys.}}
  \textbf{\bibinfo{volume}{83}}, \bibinfo{pages}{4324--4333}
  (\bibinfo{year}{1998}).

\bibitem{Das1989}
\bibinfo{author}{Das, B.} \emph{et~al.}
\newblock \bibinfo{title}{Evidence for spin splitting in
  In$_x$Ga$_{1-x}$As/In$_{0.52}$Al$_{0.48}$As heterostructures as
  B $\rightarrow$ 0}.
\newblock \emph{\bibinfo{journal}{Phys. Rev. B}} \textbf{\bibinfo{volume}{39}},
  \bibinfo{pages}{1411--1414} (\bibinfo{year}{1989}).

\bibitem{Engels1997}
\bibinfo{author}{Engels, G.}, \bibinfo{author}{Lange, J.},
  \bibinfo{author}{Sch\"apers, T.} \& \bibinfo{author}{L\"uth, H.}
\newblock \bibinfo{title}{Experimental and theoretical approach to spin
  splitting in modulation-doped In$_{x}$Ga$_{1-x}$As/InP quantum wells for
  B  $\rightarrow$ 0}.
\newblock \emph{\bibinfo{journal}{Phys. Rev. B}} \textbf{\bibinfo{volume}{55}},
  \bibinfo{pages}{R1958--R1961} (\bibinfo{year}{1997}).

\bibitem{Hu1999}
\bibinfo{author}{Hu, C.-M.} \emph{et~al.}
\newblock \bibinfo{title}{Zero-field spin splitting in an inverted
  In$_{0.53}$Ga$_{0.47}$As/In$_{0.52}$Al$_{0.48}$As heterostructure: Band
  nonparabolicity influence and the subband dependence}.
\newblock \emph{\bibinfo{journal}{Phys. Rev. B}} \textbf{\bibinfo{volume}{60}},
  \bibinfo{pages}{7736--7739} (\bibinfo{year}{1999}).

\bibitem{Pfeffer1999}
\bibinfo{author}{Pfeffer, P.} \& \bibinfo{author}{Zawadzki, W.}
\newblock \bibinfo{title}{Spin splitting of conduction subbands in
III-V heterostructures due to inversion asymmetry}.
\newblock \emph{\bibinfo{journal}{Phys. Rev. B}} \textbf{\bibinfo{volume}{59}},
  \bibinfo{pages}{R5312--R5315} (\bibinfo{year}{1999}).

\bibitem{Brosig1999}
\bibinfo{author}{Brosig, S.} \emph{et~al.}
\newblock \bibinfo{title}{Zero-field spin splitting in InAs-AlSb quantum wells
  revisited}.
\newblock \emph{\bibinfo{journal}{Phys. Rev. B}} \textbf{\bibinfo{volume}{60}},
  \bibinfo{pages}{R13989--R13992} (\bibinfo{year}{1999}).

\bibitem{Koga2002}
\bibinfo{author}{Koga, T.}, \bibinfo{author}{Nitta, J.},
  \bibinfo{author}{Akazaki, T.} \& \bibinfo{author}{Takayanagi, H.}
\newblock \bibinfo{title}{Rashba spin-orbit coupling probed by the weak
  antilocalization analysis in InAlAs/InGaAs/InAlAs quantum wells as a
  function of quantum well asymmetry}.
\newblock \emph{\bibinfo{journal}{Phys. Rev. Lett.}}
  \textbf{\bibinfo{volume}{89}}, \bibinfo{pages}{046801}
  (\bibinfo{year}{2002}).

\bibitem{Ganichev2004}
\bibinfo{author}{Ganichev, S.~D.} \emph{et~al.}
\newblock \bibinfo{title}{Experimental separation of Rashba and Dresselhaus
  spin splittings in semiconductor quantum wells}.
\newblock \emph{\bibinfo{journal}{Phys. Rev. Lett.}}
  \textbf{\bibinfo{volume}{92}}, \bibinfo{pages}{256601}
  (\bibinfo{year}{2004}).

\bibitem{Miller2003}
\bibinfo{author}{Miller, J.~B.} \emph{et~al.}
\newblock \bibinfo{title}{Gate-controlled spin-orbit quantum interference
  effects in lateral transport}.
\newblock \emph{\bibinfo{journal}{Phys. Rev. Lett.}}
  \textbf{\bibinfo{volume}{90}}, \bibinfo{pages}{076807}
  (\bibinfo{year}{2003}).

\bibitem{Grundler2000}
\bibinfo{author}{Grundler, D.}
\newblock \bibinfo{title}{Large Rashba splitting in InAs quantum wells due to
  electron wave function penetration into the barrier layers}.
\newblock \emph{\bibinfo{journal}{Phys. Rev. Lett.}}
  \textbf{\bibinfo{volume}{84}}, \bibinfo{pages}{6074--6077}
  (\bibinfo{year}{2000}).

\bibitem{Heida1998}
\bibinfo{author}{Heida, J.~P.}, \bibinfo{author}{van Wees, B.~J.},
  \bibinfo{author}{Kuipers, J.~J.}, \bibinfo{author}{Klapwijk, T.~M.} \&
  \bibinfo{author}{Borghs, G.}
\newblock \bibinfo{title}{Spin-orbit interaction in a two-dimensional electron
  gas in a InAs/AlSb quantum well with gate-controlled electron density}.
\newblock \emph{\bibinfo{journal}{Phys. Rev. B}} \textbf{\bibinfo{volume}{57}},
  \bibinfo{pages}{11911--11914} (\bibinfo{year}{1998}).

\bibitem{Matsuyama2000}
\bibinfo{author}{Matsuyama, T.}, \bibinfo{author}{K\"ursten, R.},
  \bibinfo{author}{Mei\ss{}ner, C.} \& \bibinfo{author}{Merkt, U.}
\newblock \bibinfo{title}{Rashba spin splitting in inversion layers on $p$-type
  bulk InAs}.
\newblock \emph{\bibinfo{journal}{Phys. Rev. B}} \textbf{\bibinfo{volume}{61}},
  \bibinfo{pages}{15588--15591} (\bibinfo{year}{2000}).

\bibitem{Kalevich1990}
\bibinfo{author}{Kalevich, V.} \& \bibinfo{author}{Korenev, V.}
\newblock \bibinfo{title}{Effect of electric field on the optical orientation
  of 2D electrons}.
\newblock \emph{\bibinfo{journal}{JETP Lett.}} \textbf{\bibinfo{volume}{52}},
  \bibinfo{pages}{230--235} (\bibinfo{year}{1990}).

\bibitem{KatoNature2004}
\bibinfo{author}{Kato, Y.}, \bibinfo{author}{Myers, R.~C.},
  \bibinfo{author}{Gossard, A.~C.} \& \bibinfo{author}{Awschalom, D.~D.}
\newblock \bibinfo{title}{Coherent spin manipulation without magnetic fields in
  strained semiconductors}.
\newblock \emph{\bibinfo{journal}{Nature}} \textbf{\bibinfo{volume}{427}},
  \bibinfo{pages}{50--53} (\bibinfo{year}{2004}).

\bibitem{Crooker2005}
\bibinfo{author}{Crooker, S.~A.} \& \bibinfo{author}{Smith, D.~L.}
\newblock \bibinfo{title}{Imaging spin flows in semiconductors subject to
  electric, magnetic, and strain fields}.
\newblock \emph{\bibinfo{journal}{Phys. Rev. Lett.}}
  \textbf{\bibinfo{volume}{94}}, \bibinfo{pages}{236601}
  (\bibinfo{year}{2005}).

\bibitem{KatoPRL2004}
\bibinfo{author}{Kato, Y.~K.}, \bibinfo{author}{Myers, R.~C.},
  \bibinfo{author}{Gossard, A.~C.} \& \bibinfo{author}{Awschalom, D.~D.}
\newblock \bibinfo{title}{Current-induced spin polarization in strained
  semiconductors}.
\newblock \emph{\bibinfo{journal}{Phys. Rev. Lett.}}
  \textbf{\bibinfo{volume}{93}}, \bibinfo{pages}{176601}
  (\bibinfo{year}{2004}).

\bibitem{EngelPRL2007}
\bibinfo{author}{Engel, H.-A.}, \bibinfo{author}{Rashba, E.~I.} \&
  \bibinfo{author}{Halperin, B.~I.}
\newblock \bibinfo{title}{Out-of-plane spin polarization from in-plane electric
  and magnetic fields}.
\newblock \emph{\bibinfo{journal}{Phys. Rev. Lett.}}
  \textbf{\bibinfo{volume}{98}}, \bibinfo{pages}{036602}
  (\bibinfo{year}{2007}).

\bibitem{GanichevReview2003}
\bibinfo{author}{Ganichev, S.} \& \bibinfo{author}{Prettl, W.}
\newblock \bibinfo{title}{Spin photocurrents in quantum wells}.
\newblock \emph{\bibinfo{journal}{J. Phys.: Condens. Matter}}
  \textbf{\bibinfo{volume}{15}}, \bibinfo{pages}{R935--R983}
  (\bibinfo{year}{2003}).

\bibitem{Duckheim2006}
\bibinfo{author}{Duckheim, M.} \& \bibinfo{author}{Loss, D.}
\newblock \bibinfo{title}{Electric-dipole-induced spin resonance in disordered
  semiconductors}.
\newblock \emph{\bibinfo{journal}{Nature Phys.}} \textbf{\bibinfo{volume}{2}},
  \bibinfo{pages}{195--199} (\bibinfo{year}{2006}).

\bibitem{Bloch1940}
\bibinfo{author}{Bloch, F.} \& \bibinfo{author}{Siegert, A.}
\newblock \bibinfo{title}{{Magnetic resonance for nonrotating fields}}.
\newblock \emph{\bibinfo{journal}{Phys. Rev.}}
  \textbf{\bibinfo{volume}{57}}, \bibinfo{pages}{522--527}
  (\bibinfo{year}{1940}).

\bibitem{Crooker1995}
\bibinfo{author}{Crooker, S.~A.}, \bibinfo{author}{Awschalom, D.~D.} \&
  \bibinfo{author}{Samarth, N.}
\newblock \bibinfo{title}{Time-resolved faraday rotation spectroscopy of spin
  dynamics in digital magnetic heterostructures}.
\newblock \emph{\bibinfo{journal}{IEEE J. Sel. Top. Quantum Electron.}} \textbf{\bibinfo{volume}{1}},
\bibinfo{pages}{1082--1092}
  (\bibinfo{year}{1995}).

\bibitem{MeierAPL06}
\bibinfo{author}{Meier, L.}, \bibinfo{author}{Salis, G.},
  \bibinfo{author}{Ellenberger, C.}, \bibinfo{author}{Ensslin, K.} \&
  \bibinfo{author}{Gini, E.}
\newblock \bibinfo{title}{Stray-field-induced modification of coherent spin
  dynamics}.
\newblock \emph{\bibinfo{journal}{Appl. Phys. Lett.}}
  \textbf{\bibinfo{volume}{88}}, \bibinfo{pages}{172501}
  (\bibinfo{year}{2006}).

\bibitem{Kikkawa1998}
\bibinfo{author}{Kikkawa, J.~M.} \& \bibinfo{author}{Awschalom, D.~D.}
\newblock \bibinfo{title}{Resonant spin amplification in $n$-type GaAs}.
\newblock \emph{\bibinfo{journal}{Phys. Rev. Lett.}}
  \textbf{\bibinfo{volume}{80}}, \bibinfo{pages}{4313--4316}
  (\bibinfo{year}{1998}).

\end{thebibliography}
\end{document}